\documentclass[showpacs,preprintnumbers,amsmath,amssymb]{revtex4}
\usepackage{graphicx}
\usepackage{dcolumn}
\usepackage{bm}
\begin{document}


\preprint{}


\title{Selective distillation phenomenon in two-species Bose-Einstein condensates in open boundary optical lattices\footnote{Published in Sci. Rep. 5, 17101 (2015)}}

\author{Xiao-Dong Bai,  Mei Zhang,  Jun Xiong, Guo-Jian Yang,  and  Fu-Guo Deng\footnote{Corresponding author:
fgdeng@bnu.edu.cn}}

\address{Department of Physics, Applied Optics Beijing Area Major Laboratory, Beijing normal University, Beijing 100875, China}
\date{\today }

\begin{abstract}
We investigate the formation of discrete breathers (DBs) and the
dynamics of the mixture of two-species Bose-Einstein condensates
(BECs) in open boundary optical lattices using the discrete
nonlinear Schr\"{o}dinger equations. The results show that the
coupling of intra- and interspecies interaction can lead to the
existence of pure single-species DBs and symbiotic DBs (i.e.,
two-species DBs). Furthermore, we find that there is a selective
distillation phenomenon in the dynamics of the mixture of
two-species BECs. One can selectively distil one species from the
mixture of two-species BECs and can even control dominant species
fraction by adjusting the intra- and interspecies interaction in
optical lattices. Our selective distillation mechanism may find
potential application in quantum information storage and quantum
information processing based on multi-species atoms.
\end{abstract}



\maketitle


Bose-Einstein condensates (BECs) trapped in periodic optical
potentials are an invaluable tool to study fundamental and applied
aspects of quantum optics, quantum computing, and solid state
physics\cite{IBloch2005,OMorsch2006,WSBakr2009,JSimon2011}. It is
important to understand the dynamics and transport properties of
BECs in optical lattices. One of the most interesting features of
BECs in nonlinear lattices is the existence of localized excitation,
which can propagate without changing its shape as a result of the
balance between nonlinearity and
dispersion\cite{DKCampbell2004,Campbell2004,SFlach2008,DLWang2008}.
This phenomenon is also referred to the formation of discrete
breathers (DBs). DB is an interesting discovery in nonlinear science
and has been observed in other physical systems as well, such as
micromechanical cantilever arrays\cite{MSato2006}, antiferromagnet
systems\cite{UTSchwarz1999,MSato2004}, Josephson-junction
arrays\cite{ETrias2000,AVUstinov2003}, nonlinear waveguide
arrays\cite{RMorandotti1999,HSEisenberg1998}, Tonks
gas\cite{Kolomeisky1992}, and some dissipative
systems\cite{PCMatthews2011}. In single-species BECs, many
properties of DBs have been investigated theoretically and
experimentally in the last
decade\cite{STsuchiya2008,MLarcher2011,MAntezza2007,SAubry1997,JDorignac2004,
HSakaguchi2010,MMatuszewski2005,NBoechler2010,XDBai2012}. One of its
interesting properties is that the DBs are attractors and can slow
down the relaxation processes in dissipative
systems\cite{RLivi2006,HHennig2013,GSNg2009}. Moreover,  some
studies\cite{GSNg2009,BRumpf2004,HHennig2010} on the collision of a
stationary DB with a lattice excitation (a moving breather or
phonon) show that if the amplitude of the lattice excitation is
small, it will be reflected entirely from the DB, while with the
amplitude beyond a specific threshold, a part of the incident atoms
transmit through the DB.

The previous works are mainly focused on single-species BECs.
Actually, the two- and multi-species BECs have been observed in
experiment and attracted much attention. In 2008, Thalhammer
\emph{et al.}\cite{GThalhammer2008}  observed an interesting mixture
of heteronuclear BECs in experiment,  where $\ ^{41}$K and $\
^{87}$Rb atoms are condensed together in an optical lattice. An
important property of this mixture is that the interspecies
scattering length  describing the effective colliding interaction
between  $\ ^{41}$K and $\ ^{87}$Rb atoms can be tuned over a wide
(both positive and negative) range using a magnetic Feshbach
resonance, and their own intraspecies scattering length remains
positive for each species. That is, in this mixture both the inter-
and intraspecies interactions can be varied and controlled
completely. Subsequently, both the stable mixture\cite{SSugawa2011}
of the isotopes $\ ^{168}$Yb and $\ ^{174}$Yb, and the unstable
mixture\cite{TFukuhara2009} of the isotopes $\ ^{174}$Yb and $\
^{176}$Yb were  obtained. In 2011, two-species BECs have been
realized\cite{YJLin2011}  with the mixture of  two hyperfine states
of $\ ^{87}$Rb, which is spin-orbit-coupled (SO-coupled) BECs.

Recently, some interesting physical phenomena and unique properties
have been discovered in multi-species BECs in optical lattices, such
as multi-species gap solitons in spinor
BECs\cite{BJDabrowskaWuster2007}, dark-dark solitons and
modulational instability in miscible two-species
BECs\cite{MAHoefer2011}, unstaggered-staggered
solitons\cite{BAMalomed2012}, and the other two-species
solitons\cite{DSWang2010,AGubeskys2006,AGubeskys2008,
SkGolamAli2009,HACruz2007,FKhAbdullaev2008,ZShi2008,
AIYakimenko2012,VMPerezGarcia2005} in two-species BECs. Also, it has
been found that  the mixing with the second atomic species can lead
to some different physical
phenomena\cite{JRuostekoski2007,RCampbell2015}. For example, the
interspecies interaction of the two-species BECs can result in the
phase separation in a harmonic trap,  i.e.,  the two species may be
immiscible\cite{POhberg1998,MMatuszewski2007}.  In 2008, Papp
\emph{et al.}\cite{SBPapp2008} found  that the repulsive interaction
between atoms of different species can leave the mixture of
two-species BECs far from its ground  state in experiment.

In this paper, we  numerically investigate  the formation of DBs in
two-species BECs inside open optical lattices using the discrete
nonlinear Schr\"{o}dinger equations (DNLSEs). We find that the
coupling of intra- and interspecies interaction can lead to the
existence of pure single-species DB and symbiotic DBs, i.e., the DBs
of species $1$ and $2$  locate together with the same or different
species fraction at the same sites in  open optical lattices.
Furthermore, we explore  the dynamics of the mixture in two-species
BECs with a pure single-species DB  in open optical lattices.
Interestingly, we find that there is a selective distillation
phenomenon in both the mixture of  initial condition selected
randomly and that of symbiotic DB. That is, by adjusting the
interspecies interaction one can make  one species transmit through
the DB and the other be blocked, therefore increasing  the relative
proportion of the ultracold atoms in the former. Moreover, one can
also improve  the dominant specie fraction of the mixture of
two-species BECs by tuning the interspecies interaction in three
mixtures: initial condition selected randomly, moving symbiotic DBs,
and stable symbiotic DBs. This phenomenon is potentially  useful in
quantum information storage and quantum information processing based
on multi-species atoms.

\begin{figure}[!htp]  
\begin{center}
\includegraphics[width=18cm]{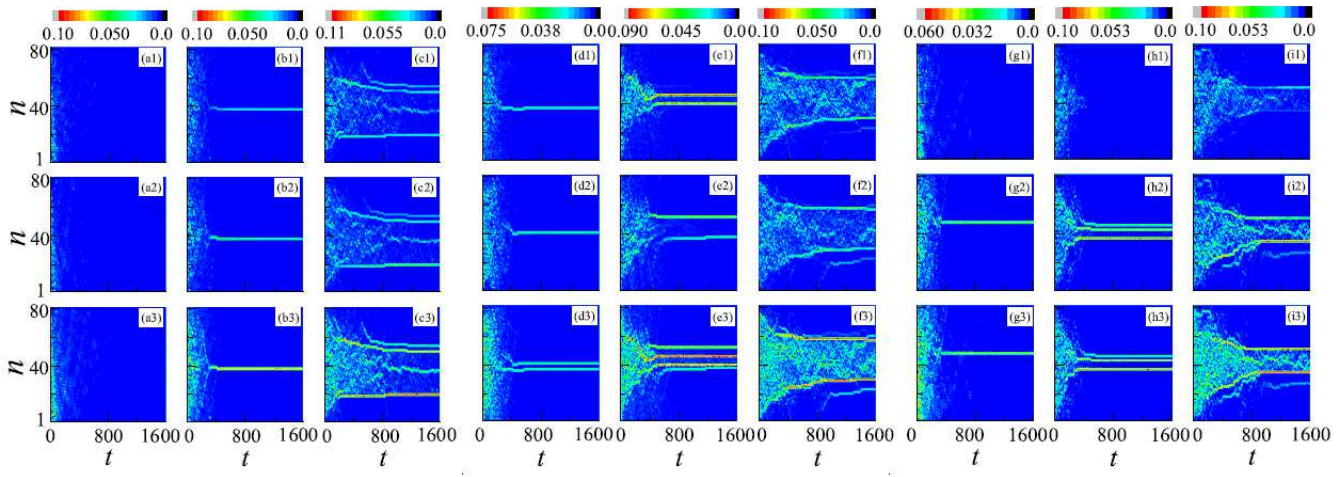}
\end{center}
\caption{ \label{Fig1}   Formation of DBs of two-species BECs in
open optical lattices with $M=81$ sites. The color code shows
$|\psi_{1,n}|^2$ (normalized to 1 at $t=0$ for species 1),
$|\psi_{2,n}|^2$ (normalized to 1 at $t=0$ for species 2), and
$|\psi_{1,n}|^2+|\psi_{2,n}|^2$ in the first, the second, and the
third rows, respectively.  The initial condition is a homogeneously
populated lattice with random phases at each site randomly drawn
from $[0,2\pi]$. The boundary dissipation rates at sites $1$ and $M$
are $ \gamma_1= \gamma_2= 0.3$. The other parameters are chosen as
follows: $\Lambda_{1,2}=0.1 < \Lambda_b$ for (a1)-(a3), (d1)-(d3),
and (g1)-(g3); $\Lambda_{1,2}=0.5>\Lambda_b$ for (b1)-(b3),
(e1)-(e3), and (h1)-(h3), and $\Lambda_{1,2}=1>\Lambda_b$ for
(c1)-(c3), (f1)-(f3), and (i1)-(i3). (a)-(c):
$\Lambda_{1,1}=\Lambda_{2,2}=0.1$; (d)-(f):
$\Lambda_{1,1}=\Lambda_{2,2}=0.6$; (g)-(i): $\Lambda_{1,1}=0.1$ and
$\Lambda_{2,2}=0.6$.}
\end{figure}

\section*{Results}
\subsection{The model of two-species Bose-Einstein condensates.}
Two-species BECs can be created by simultaneously confining
different atomic species in the same magnetic trap, including those
of two different kinds of atoms and those of the same atoms in two
different hyperfine states. For instance, a strongly repelling
two-species system of different species can be created using $\
^{41}$K-$\ ^{87}$Rb atoms in an optical
lattice\cite{GThalhammer2008}. Another two-species BECs  were
experimentally realized in hyperfine spin states of $\ ^{87}$Rb,
$|\uparrow\rangle\equiv|F=1,m_f=0\rangle$ and
$|\downarrow\rangle\equiv|F=1,m_f=-1\rangle$, which is called
SO-coupled BECs and resulted from a pair of counterpropagating Raman
beams coupling the atomic states\cite{YJLin2011} $|\uparrow\rangle$
and $|\downarrow\rangle$. In these systems, the inter- and
intraspecies interaction strengths can be controlled by a magnetic
Feshbach resonance or adjusting the angle between the Raman beams.

We start our study with the following coupled Gross-Pitaevskii
equations (GPEs) describing the dynamics of the two-species
BECs\cite{DSHall1998,ASinatra1999,YEto2015}
\begin{eqnarray}\label{eq1}
i\hbar\frac{\partial \Psi _j(\overset{\to }{r})}{\partial
t}\;=\;\left(\!\!-\frac{\hbar^2 }{2m_j}\triangledown
^2\!+\!V_j(\overset{\to }{r})+\!\!\!\sum _{i=1,2}\!\!
g_{i,j}|\Psi_{i}(\overset{\to }r)|^2\!\!\right)\!\!\Psi
_j(\overset{\to }{r}).
\end{eqnarray}
Here $\Psi_{j}(\overset{\to }r)$ denotes the condensate for species
$j$ ($=1,2$). The coefficient $g_{i,j}$ represents the interaction
between two  atoms from  species $i$ and  $j$, which is defined as
\begin{eqnarray}\label{eq2}
g_{i,j}\;=\;\frac{2\pi \hbar^2 a_{i,j}}{\mu _{i,j}},
\end{eqnarray}
where $a_{i,j}$ is the intraspecies ($i=j$) or interspecies ($i\neq
j$) scattering lengths and $\mu_{i,j}=m_im_j/(m_i+m_j)$ is the
reduced mass of the atomic pair. The external potential
$V_j(\overset{\to }{r})$ is generally a superposition of a harmonic
trapping potential $V_{H,j}(\overset{\to }{r})$ and the periodic
optical lattice potential $V_{L,j}(\overset{\to }{r})$, that is,
\begin{eqnarray}
V_j(\overset{\to }{r})\;=\;V_{H,j}(\overset{\to
}{r})+V_{L,j}(\overset{\to }{r}).
\end{eqnarray}
Here
\begin{eqnarray}\label{eq3}
V_{H,j}(\overset{\to }{r})&=&\frac{1}{2}m\left(\omega
_{jx}^2x^2+\omega _{jy}^2y^2+\omega _{jz}^2z^2\right),\\
V_{L,j}(\overset{\to }{r})&=&V_{0,j}\sin ^2\left(\pi x/a+\varphi
_j\right),
\end{eqnarray}
where $a$ is the lattice spacing.  When the lattices are
sufficiently deep, one can work in the tight-binding limit, and the
condensate is well localized around potential minima. The condensate
parameter can be written as
\begin{eqnarray}\label{eqstate}
\Psi_j(t)\;=\;\sqrt{N_j}\sum_n \psi _{j,n}(t)\phi
_{j,n}(\overset{\to }{r}),
\end{eqnarray}
where $n$ $(=1,\cdots,M)$ is the index of the site and $\phi
_{j,n}(\overset{\to }{r})$ accounts for the ground state of the
correspondingly isolated $n$-th potential.  $M$ is the number of
lattice sites. $|\psi _{j,n}(t)|^2$ represents the number of the
$j$-th species atoms at the $n$-th lattice site. By inserting Eq.
(\ref{eqstate}) into Eq. (\ref{eq1}) and integrating out the spatial
degree  of freedom, one can obtain
DNLSEs\cite{ATrombettoni2001,UShrestha2012}:
\begin{eqnarray}\label{eq5}
\!i\frac{\partial \psi _{j,n}} {\partial
t}\!&=&\!\left(\lambda_{j,j} \left|\psi_{j,n}\right|^{2}+
\lambda_{1,2}\left|\psi_{3-j,n}\right|^{2}
\right)\psi_{j,n}-\frac{1}{2}\left(\psi _{j,n-1}+\psi
_{j,n+1}\right).
\end{eqnarray}
The total atomic population inside the optical lattice for each of
the species $1$ and $2$ is $\sum_n |\psi_{j,n}(t=0)|^2=N_j(t=0)$.
Here the atomic distribution of each species over the entire lattice
is normalized to unity: $N_j(t=0)=1$. The wave functions can be
assumed as $\psi_{j,n}=A_{j,n}\exp(i\theta_{j,n})$, where $A_{j,n}$
and $\theta_{j,n}$ represent the amplitudes and the phases of
species $j$ at site $n$, respectively. In Eq. (\ref{eq5}), the time
has been re-scaled $t\rightarrow[\hbar/(2J)]t$ with the assumption
$J=J_1=J_2=-\int d^3 \overset{\to
}{r}\left(\frac{\hbar^2}{2m_j}\nabla \phi_{j,n}\nabla
\phi_{j,n+1}+V(\overset{\to }{r})\nabla \phi_{j,n}\nabla
\phi_{j,n+1}\right)$ being the tunneling rate between the
nearest-neighbor sites. The intra- and interspecies interactions are
described by the parameters $\lambda_{j,j}=\frac{g_{j,j}}{2J}\int
d^{\, 3}\overset{\to }{r}|\phi_{j,n}|^4$ and
$\lambda_{1,2}=\frac{g_{1,2}}{2J}\int d^{\,3}\overset{\to
}{r}|\phi_{1,n}|^2|\phi_{2,n}|^2$, respectively. Here, we focus on
the repulsive cases, i.e., $\lambda_{j,j} > 0$ and
$\lambda_{1,2}>0$.

\subsection{Formation of discrete breathers in two-species Bose-Einstein condensates.} The formation of DBs in two-species BECs ($N_1=N_2$) in open optical
lattices can be investigated systematically by introducing the
initial effective mean-field intra- and interspecies  interactions
per site as\cite{GSNg2009,HHennig2013}
\begin{eqnarray}\label{eq8}
\Lambda_{j,j}\;=\;\frac{\lambda_{j,j}N_{j,j}(t=0)}{M}, \;\;\;\;\;
\Lambda_{1,2}\;=\;\frac{\lambda_{1,2}\sqrt{N_{1}N_{2}}(t=0)}{M}.\;\;\;\;
\end{eqnarray}
It has been demonstrated that in single-species BECs if the initial
effective mean-field interaction $\Lambda$ is larger than a critical
value $\Lambda^*\approx0.472$ (which is gained from Eq. (17) of Ref.
~\cite{HHennig2013} when $M=81$), because of self-localization
mechanism, the stable DBs can be created in open optical lattices
with boundary dissipation\cite{GSNg2009,HHennig2013}. If $\Lambda$
is less than $\Lambda^*$, DBs cannot be formed and atoms will decay.
In two-species BECs, the DBs are still rooted from self-localization
mechanism. If $\Lambda_{1,2}=0$, species $1$ and $2$ are independent
and one can get the corresponding critical values
$\Lambda_{1,1}^*=\Lambda_{2,2}^*\approx0.472$. If
$\Lambda_{1,2}\neq0$, species $1$ and $2$ are dependent, and for the
special case $\Lambda_{1,1}=\Lambda_{2,2}=0$,  one can analogize the
BECs with species $1$ and $2$ to  single-species BECs and its
corresponding critical value is $\Lambda_{1,2}^*\approx0.472$. Here
and after, we assume the critical value is $\Lambda_b=0.472$.
According to the difference of the parameters $\Lambda_{1,1}$ and
$\Lambda_{2,2}$, the following investigation should be divided into
three cases: (I) $\Lambda_{1,1}=\Lambda_{2,2}=0.1<\Lambda_b$; (II)
$\Lambda_{1,1}=\Lambda_{2,2}=0.6>\Lambda_b$; (III)
$\Lambda_{1,1}=0.1<\Lambda_b,$ and $\Lambda_{2,2}=0.6>\Lambda_b$. In
these three cases, the interspecies interaction $\Lambda_{1,2}$ will
paly an important role and will impact strongly on the formation and
dynamics of DBs for two-species BECs in open optical lattices.

Let us consider the dissipation case with   atoms initially
distributed uniformly  at each site with random phases
$\theta_{j,n}$, which  reads
\begin{eqnarray}\label{eq9}
\psi_{j,n}(t=0)\;=\;A_{j,n}e^{i\theta_{j,n}}\;=\;\frac{1}{\sqrt{M}}e^{i\theta_{j,n}}.
\end{eqnarray}
Here $\theta_{j,n}\in[0,2\pi]$ is an arbitrary value. We assume that
the boundary dissipation rates of lattices  at sites 1 and M are
$\gamma_1=\gamma_2=0.3$ for both species.

The formation of DB in two-species BECs in open optical lattices
with $M=81$ for the three cases (I), (II), and (III) with
 different $\Lambda_{1,2}$ are shown in Fig.
\ref{Fig1}. In the first row, the color code shows $|\psi_{1,n}|^2$,
which is normalized to $1$ at $t=0$ and describes the density of
species $1$. In the second and the third rows, the color codes show
$|\psi_{2,n}|^2$ which is also normalized to $1$ at $t=0$ and
describes the density of species $2$, and
$|\psi_{1,n}|^2+|\psi_{2,n}|^2$ describing the sum density of both
species $1$ and $2$, respectively.

In Fig. \ref{Fig1}(a1)-(a3),
$\Lambda_{1,1}=\Lambda_{2,2}=\Lambda_{1,2}=0.1< \Lambda_b$ is very
small, one can see that there is no DB in both species $1$ and $2$,
and the atoms in the lattices get dissipated. That is, when both
intra- and interspecies interactions  are smaller than their
critical values, no DBs can be formed. In Fig. \ref{Fig1}(b1)-(b3),
$\Lambda_{1,2}=0.5$ is a bit larger  than the critical value
$\Lambda_b$, and one DB can be formed for each of species $1$ and
$2$ and they are located in the same position, as shown in Fig.
\ref{Fig1}(b3). It can be called a symbiotic DB. In Fig.
\ref{Fig1}(c1)-(c3) with $\Lambda_{1,2}=1>\Lambda_b$,  a few DBs can
be formed for each of species $1$ and $2$, and the locations of DBs
for species $1$ are the same as those for species $2$. That is, if
$\Lambda_{1,2}>\Lambda_b$ and $\Lambda_{1,1}=\Lambda_{2,2}<
\Lambda_b$, the properties of the two-species BECs are determined
nearly by interspecies interactions $\Lambda_{1,2}$, not the
intraspecies interactions $\Lambda_{1,1}$ and $\Lambda_{2,2}$. Under
this condition, the mixture of species $1$ and $2$ are analogous to
single-species, and hence its properties are similar to those of
single-species BECs. Thus, one can see that the dynamical properties
of species $1$ and $2$ are exactly the same, and the DBs of the two
species always co-exist in the same positions, as shown in Fig.
\ref{Fig1}(b2)-(c3).

In Fig. \ref{Fig1}(d1)-(d3),
$\Lambda_{1,1}=\Lambda_{2,2}=0.6>\Lambda_b$ and
$\Lambda_{1,2}=0.1<\Lambda_b$. It is seen that the properties of the
species $1$ and $2$ are determined by their intraspecies
interactions $\Lambda_{1,1}$ and $\Lambda_{2,2}$, but not
interspecies interactions $\Lambda_{1,2}$. At this time, species $1$
and $2$ can be considered as two non-interacting species. Thus, one
can see that one DB is formed in each of species $1$ and $2$.
Different from those in Fig. \ref{Fig1}(b1)-(b3), the two DBs of
species $1$ and $2$ are not located at the same positions, as shown
in Fig. \ref{Fig1}(d3). In Fig. \ref{Fig1}(e1)-(e3),
$\Lambda_{1,2}=0.5$ is larger a little than the critical value
$\Lambda_b$, and  two DBs can be formed in each of species $1$ and
$2$ and they are located at the different positions, as shown in
\ref{Fig1}(e3). That is, the formation process of DBs in species $1$
and $2$ are still independent. In both Fig. \ref{Fig1}(f1)-(f3)
$\Lambda_{1,2}=1>\Lambda_b$, species $1$ and $2$ cannot be
considered as the two independent species, and two or more strong
DBs can be formed for each of species $1$ and $2$ and they co-exist
in the same positions. Moreover, when $\Lambda_{1,2}$ is large
enough, the DB composed of only species $1$ or $2$ can prevent atoms
of both species from dissipating out of the lattices, as shown in
Fig. \ref{Fig1}(f1)-(f3).

In Fig. \ref{Fig1}(g1)-(g3) ($\Lambda_{1,2}=0.1$) and (h1)-(h3)
($\Lambda_{1,2}=0.5$), the formation processes of DBs in species $1$
are determined by the interspecies interactions $\Lambda_{1,2}$,
while those in species $2$ are determined by both the interspecies
interactions $\Lambda_{1,2}$ and intraspecies interactions
$\Lambda_{2,2}$. Thus, one can see that no DB is formed in species
$1$, as shown in  Fig. \ref{Fig1}(g1) and (h1), and one or more DBs
can be formed in species $2$, as shown in Fig. \ref{Fig1}(g2) and
(h2). In Fig. \ref{Fig1}(i1)-(i3) with $\Lambda_{1,2}=1>\Lambda_b$,
due to the interplay of species $2$, DBs can be formed not only in
species $2$ but also in species $1$. The DBs of species $1$ are
weaker than those of species $2$. That is, the formation of DBs of
the two-species BECs is dominated by species $2$, and the DBs of
species $1$ are like appurtenances.

\section*{Selective distillation of ultracold atomic gas}
From Figs. \ref{Fig1}(f1)-(f3) and (i1)-(i3), one can see that the
DBs of both species $1$ and $2$ can prevent the atoms from
transferring through them when the interspecies interactions
$\Lambda_{1,2}$ are large enough. However, Fig. \ref{Fig1}(a1)-(a3)
predicts that species $1$ and $2$ are independent of each other when
the interspecies interaction  $\Lambda_{1,2}$ is small or even
vanish, that is, the DB of species $j$   has  an important effect on
the transfer process of its own but not on the other species $3-j$.
With this interesting mechanism, we propose a theoretical scheme to
selectively distill one species from the mixture of two-species BECs
in optical lattices. In order to describe the principle of our
scheme clearly, let us define two new parameters,
\begin{eqnarray}\label{eq10}
Rd_{j}(t)\;=\;\frac{\sum_{n=k_1}^{k_2}|\psi_{j,n}(t)|^2
}{\sum_{n=k_1}^{k_2}|\psi_{1,n}(t)|^2 +\sum
_{n=k_1}^{k_2}|\psi_{2,n}(t)|^2 }
\end{eqnarray}
and
\begin{eqnarray}\label{Rp}
Rp_{j}(t)\;=\;\frac{\sum_{n=k_1}^{k_2}|\psi_{j,n}(t)|^2
}{\sum_{n=1}^{M}\left(|\psi_{1,n}(t=0)|^2
+|\psi_{2,n}(t=0)|^2\right) }.
\end{eqnarray}
$Rd_{j}(t)$ describes the dominant species fraction of species $j$
in the mixture of two-species BECs from the sites $k_1$ to $k_2$ in
the optical lattice at time $t$, while $Rp_{j}(t)$ describes the
relative proportion of the atoms in species $j$ from the sites $k_1$
to $k_2$ in the mixture to all the atoms in the entire optical
lattice. Here our investigation is mainly focused on two typical
cases, i.e.,  initial condition selected randomly and moving
symbiotic DB. It is worth noticing that we focus on the impact of DB
on the dynamics of BECs, where the DB mainly occupies three sites
and the effective interaction is $\Lambda_{i,j}=\lambda_{i,j}/3$.
For convenience, we choose $\lambda_{i,j}$ to describe the dynamics
of two-species BECs below, and this parameter has been used in
previous works\cite{HHennig2010,XDBai2015}.

\begin{figure}[!htp] 
\includegraphics[width=18cm]{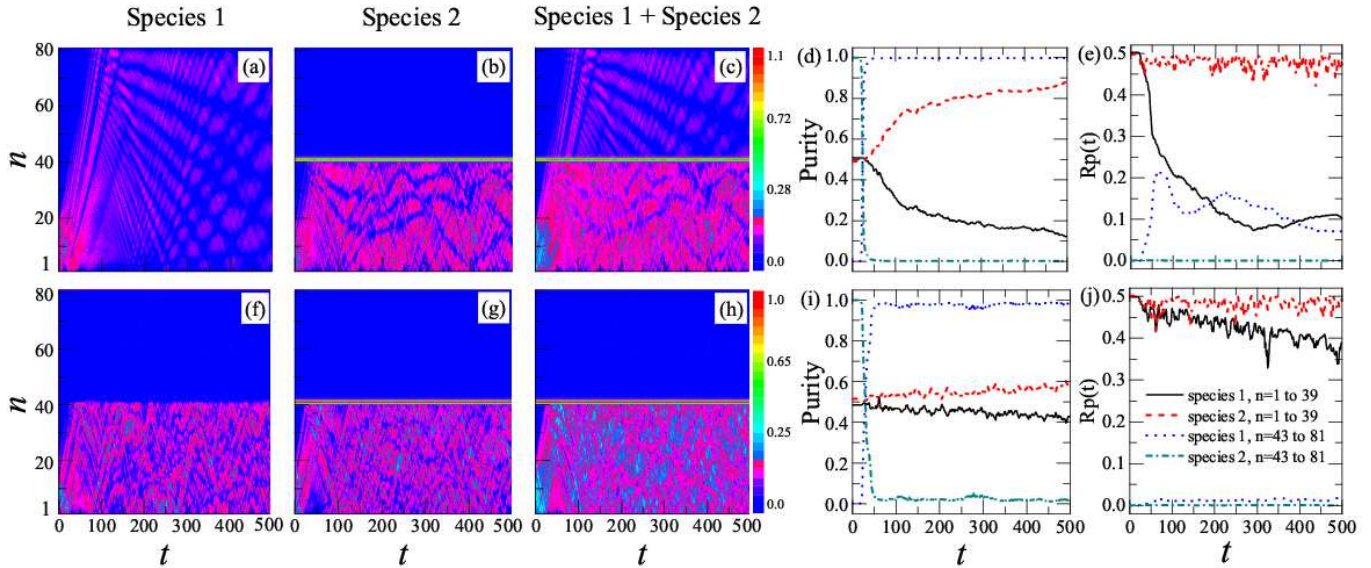}
\caption{\label{Fig4}  Selective distillation phenomenon in the
chaos mixture of two-species BECs. The initial conditions are
presented in Eqs. (\ref{eq11}) and (\ref{eq12}). The color codes
show $|\psi_{1,n}|^2$, $|\psi_{2,n}|^2$, and
$|\psi_{1,n}|^2+|\psi_{2,n}|^2$ in the first, the second, and the
third columns, respectively. In the fourth and fifth columns, the
solid and dashed lines represent the $Rd_j(t)$ and the $Rp_j(t)$ of
species $1$ and $2$ at sites ranging from $1$ to $39$, respectively.
The dotted and dot-dashed lines represent the $Rd_j(t)$ and the
relative proportions of species $1$ and $2$ at sites ranging from
$43$ to $81$, respectively. They are obtained from Eqs. (\ref{eq10})
and (\ref{Rp}). In all cases, $\lambda_{1,1}=\lambda_{2,2}=4$,
$\lambda_{1,2}=0$ in (a)-(e), and $\lambda_{1,2}=3$ in (f)-(j).}
\end{figure}

\subsection{Selective distillation phenomenon in the chaos mixture of BECs.}

Let us assume that the initial condition is chaos  with random
amplitudes and phases at sites  $1$ to $20$ in optical lattices with
$M=81$, and the other sites are empty (that is,  their amplitudes
are zero). There is a pure DB  of species $2$ at the middle site of
optical lattice. That is, the initial condition reads as
\begin{eqnarray}\label{eq11}
&& \psi_{j,n}(t=0)=A_{j,n}e^{i\theta_{j,n}}, \;\;\;\;\;n=1\ldots20  \nonumber\\
&&A_{j,n}\in[0,0.4],\;\;\;\;\;\;\;\theta_{j,n}\in[0,2\pi],\\
&&A_{j,n}=0, \;\;\;\;\;\;\;\;\;\;\;\;\;\;\;\;\;\;\;\;\texttt{else},\nonumber
\end{eqnarray}
and
\begin{eqnarray}\label{eq12}
\begin{array}{c}
\!\!\!\!A_{2,40}(t=0)=A_{2,42}(t=0)=0.1667,\;\;\;\;\;\;\;\;\;\;\;\;\;\; \\A_{2,41}(t\!=\!0)\!=\!\left(1\!-\!A_{2,40}^2(t\!=\!0)\!-\!A_{2,42}^2(t\!=\!0)\right)^{1/2}. \\
\end{array}
\end{eqnarray}
Eq. (\ref{eq11}) represents the initial condition selected randomly,
and $A_{j,n}$ and $\theta_{j,n}$ are arbitrary values.  Eq.
(\ref{eq12}) represents the DB of species $2$.

Using the initial condition and Eqs. (8) and (9) with
$\gamma_1=\gamma_2=0$, we simulate numerically the dynamics of
two-species BECs, shown in Fig. \ref{Fig4} with
$\lambda_{1,1}=\lambda_{2,2}=4$. The color codes show
$|\psi_{1,n}|^2$, $|\psi_{2,n}|^2$, and
$|\psi_{1,n}|^2+|\psi_{2,n}|^2$ in the first, the second, and the
third columns, respectively. In the fourth and fifth columns, the
different lines represent the $Rd_j(t)$ and $Rp_j(t)$ for species
$1$ and $2$ at the different sites ranging from $1$ to $39$ or from
$43$ to $81$, respectively.

In Fig. \ref{Fig4}(a)-(e), $\lambda_{1,2}=0$,  which means these two
species are independent of  each other. If the DB is composed of
only species $2$, it has effect only on the dynamics of species $2$,
but not  on that of species $1$. As shown in Fig. \ref{Fig4}(a)-(c),
the part composed of species $1$ in chaos can transmit  through the
DB without any hindrance, but the part composed of species $2$ in
chaos has been blocked  by the DB. The change of the  $Rd_j(t)$ of
the two species is shown in Fig. \ref{Fig4}(d). When $t=0$,
$Rd_j(t=0)$ at the sites $1-39$ are $50\%$, shown with the solid and
dashed lines, respectively. At sites   $43-81$, since a small number
of atoms of the DB can move to this area, there is only species $2$,
no species $1$, that is, the $Rd_j(t)$ of species $1$ and $2$ are
$0$ and $1$, respectively, shown with the dotted and dash-dotted
lines in Fig. \ref{Fig4}(d) at the beginning ($t\sim0$),
respectively. Subsequently, the $Rd_j(t)$   change   with  time $t$.
After $24$ time steps, the chaos extends to the DB, and then a part
of species $1$ transmits through it. At  sites  $43-81$,  the
$Rd(t)$ of species $1$ first increases and then approaches the
stable value of nearly $100\%$, but that of species $2$ is nearly
$0$. In other words, species $1$ with a high $Rd_1(t)$ is extracted
from the chaos at the sites $1-39$ when $\lambda_{1,2}$ is far
smaller than the critical value. We call this phenomenon the
distillation of ultracold atomic gas.

To show this distillation phenomenon explicitly, we calculate the
$Rp_j(t)$ of species $1$ and $2$ at the two different areas (one is
composed of the sites from 1 to 39 and the other from 43 to 81),
shown in Fig. \ref{Fig4}(e). When $t=0$, $Rp_j(t=0)$ of species $1$
and $2$ at  sites  $1-39$ in the mixture are $50\%$, shown with the
solid and dashed lines, respectively,  and $Rp_j(t)$ at sites
$43-81$ are $0$, shown with the dotted and dash-dotted lines,
respectively. After $24$ time steps, $Rp_2(t)$ at sites $1-39$  and
$43-81$ are constant, and $Rp_1(t)$ at sites $1-39$ decrease and
that at sites $43-81$ can increase to $21.6\%$. That is, at most
$21.6\%$ of species $1$ can be distilled from the mixture, and
$Rp_1(t)$ at sites $43-81$ (the dotted line) can represent the
efficiency of this distillation.

When $\lambda_{1,2}=3$,  one can see that both species $1$ and $2$
are prevented by the DB and cannot transmit  through it, as shown in
Fig. \ref{Fig4}(f)-(h). In Fig. \ref{Fig4}(i), the $Rd_j(t)$ of
species $1$ and $2$ at  sites   $43-81$ are also nearly $100\%$ and
$0$ at last, respectively. However, $Rp_1(t)$ at sites $43-81$ is
about $1\%$, that is, the number of  atoms that transmit through the
DB is very small  due to the blocking of the DB, and species $1$ is
not effectively distilled. Thus, the DB plays a role of inhibitting
the transmission of both species $1$ and $2$ when $\lambda_{1,2}$ is
much larger than the critical value.

From the discussion above, one can see that the distillation for the
chaos depends strongly on the species of DB and the value of
$\lambda_{1,2}$. The former decides which species will be distilled,
and the latter can control the efficiency of distillation.
Therefore, by adjusting the interspecies interaction
$\lambda_{1,2}$,  one can make one species transmit through the DB
and the other be blocked, and increase the dominant species fraction
of the ultracold atoms in the former. We call this the selective
distillation phenomenon for the ultracold atoms in the mixture of
two-species BECs.

\begin{figure}[!htp] 
\includegraphics[width=18cm]{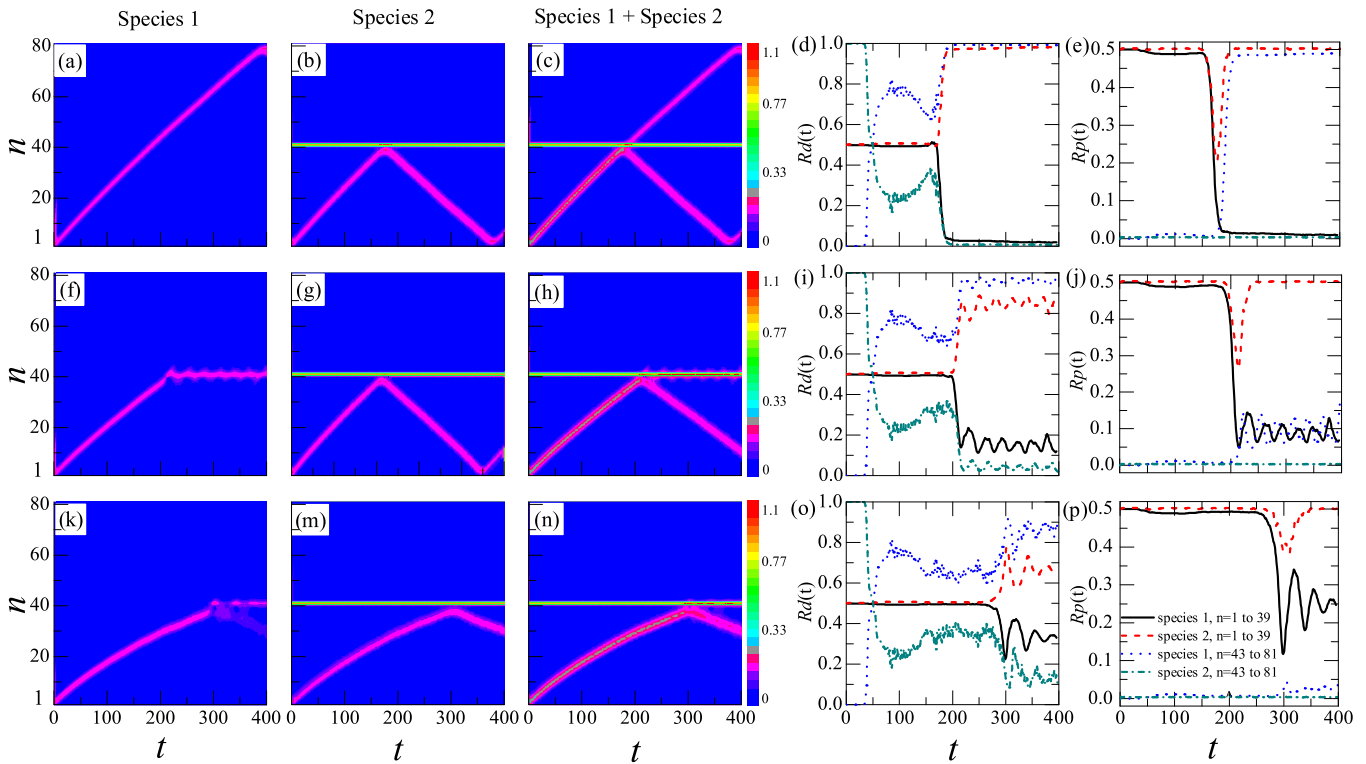}
\caption{\label{Fig5}   Selective distillation phenomenon of a
moving symbiotic DB in two-species BECs. The initial conditions are
presented in Eqs. (\ref{eq12}) and (\ref{eq13}). The color codes
show $|\psi_{1,n}|^2$, $|\psi_{2,n}|^2$, and
$|\psi_{1,n}|^2+|\psi_{2,n}|^2$ in the first, the second, and the
third columns, respectively. In the fourth and fifth columns, the
solid and dashed lines represent the $Rd_j(t)$ and $Rp_j(t)$ of
species $1$ and $2$ at sites ranging from $1$ to $39$, respectively.
The dotted and dot-dashed lines represent the $Rd_j(t)$ and
$Rp_j(t)$ of species $1$ and $2$ at sites ranging from $43$ to $81$,
respectively. They are obtained from Eqs. (\ref{eq10}) and
(\ref{Rp}). In all cases, $\lambda_{1,1}=\lambda_{2,2}=4$,
$\Lambda_{1,2}=0$ in (a)-(e), $\lambda_{1,2}=0.2$ in (f)-(j), and
$\lambda_{1,2}=0.35$ in (k)-(p).}
\end{figure}

\subsection{Selective distillation phenomenon in the mixture of moving symbiotic DB.}
Let us assume that a symbiotic DB of  two-species BECs moves to a
pure DB composed of species $2$ which is represented by Eq.
(\ref{eq12}). The initial condition for the moving  symbiotic DB
reads as
\begin{eqnarray}\label{eq13}
\begin{array}{c}
 A_{j,1}(t=0)=A_{j,4}(t=0)=0.2, \\A_{j,2}(t=0)=A_{j,3}(t=0)=0.33. \\
\end{array}
\end{eqnarray}
The dynamics of the moving DB under DNLSE with three given
interspecies interactions $\lambda_{1,2}$ is shown in Fig.
\ref{Fig5} with $\gamma_1=\gamma_2=0$ and
$\lambda_{1,1}=\lambda_{2,2}=4$.

In Fig. \ref{Fig5}(a)-(e), $\lambda_{1,2}=0$. When the symbiotic DB
moves to the stable DB, the part composed of species $1$ of this
moving DB can transmit through the stable DB without any hindrance,
but that of species $2$ is reflected by the stable DB, as shown in
Fig. \ref{Fig5}(a)-(c). The $Rd_j(t)$  and $Rp_j(t)$  of these two
species are shown in Fig. \ref{Fig5}(d) and (e), respectively.  When
$t=0$, $Rd_j(t)$  and $Rp_j(t)$ of the two species in this symbiotic
DB are nearly $50\%$, as shown in Fig. \ref{Fig5}(d)-(e) with the
solid and dashed lines, respectively.  After $169$ time steps,  this
symbiotic DB arrives at the stable DB and then collides with it.
From Fig. \ref{Fig5}(c), one can see that the transmitted part is
composed of species $1$. Accordingly, $Rd(t)$ and $Rp(t)$ of species
$1$ at sites $43-81$ is increased to nearly  $100\%$ and $48.8\%$,
shown in Fig. \ref{Fig5}(d) and (e) with the dotted lines,
respectively. That is,  species $1$ is distilled from the symbiotic
DB with a high efficiency.

When $\lambda_{1,2}=0.2$,  the dynamics of  the moving symbiotic DB
is shown in Fig.\ref{Fig5} (f)-(j).  Different from  Fig.
\ref{Fig5}(a)-(e), the part composed of species $1$ of this
symbiotic DB does not transmit through the stable DB but mixes with
it, and the part composed of species $2$ is reflected, as shown in
Fig. \ref{Fig5}(f)-(h). Their $Rd_j(t)$  and $Rp_j(t)$ are presented
in Fig. \ref{Fig5}(i) and (j), respectively. One can find that the
$Rd(t)$ of species $1$ at  sites $43-81$ increases suddenly to a
relatively stable value at time $t=200$. During this time, $Rp(t)$
of species $1$  at sites $43-81$ increases to $10\%$.

When $\lambda_{1,2}=0.35$, the dynamics of  the moving symbiotic DB
is shown in Fig. \ref{Fig5}(k)-(p). In the part composed of species
$1$ of this DB, some is mixed with the stable DB and the other part
is reflected,  as shown  in Fig. \ref{Fig5}(k), which is a new
phenomenon and not yet understood so far.  The part composed of
species $2$ of this moving DB is reflected by the stable DB. Their
$Rd_j(t)$  and $Rp_j(t)$  are presented in Fig. \ref{Fig5}(o) and
(m), respectively.  One can find that the $Rd(t)$ of the species $1$
at  sites $43-81$ is lower than that in Fig. \ref{Fig5}(d)-(i), and
$Rp_1(t)$ is $2\%$, shown with dotted lines  in Fig. \ref{Fig5}(m).

The dynamics of the symbiotic DB shows that there is also a
selective distillation phenomenon in the transport for the moving
atoms in two-species BECs with a DB. One can selectively distil one
species from the moving symbiotic DB and control its $Rd_j(t)$ by
adjusting the interspecies interaction $\lambda_{1,2}$.

Certainly, there  exist some atoms including species $1$ and $2$
escaping out of the symbiotic DB and spreading freely to the stable
DB. Since the stable DB can prevent the atoms from species $2$ but
not $1$, after $47$ time steps, the $Rd_j(t)$ of species $1$ and $2$
change suddenly, shown in Fig. \ref{Fig5}(d), (i), and (o).

\begin{figure}[!htp] 
\begin{center}
\includegraphics[width=8.5cm]{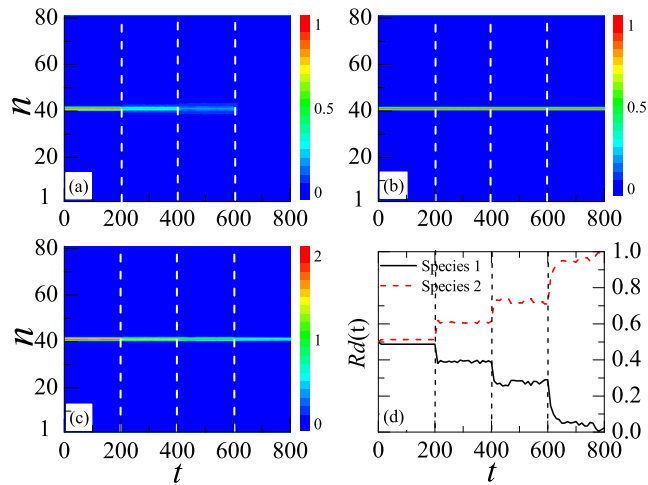}
\end{center}
\caption{\label{Fig6}   Selective distillation phenomenon of a
stable symbiotic DB in two-species BECs. The color codes show
$|\psi_{1,n}|^2$ (normalized to $1$ at $t=0$ for species $1$),
$|\psi_{2,n}|^2$ (normalized to $1$ at $t=0$ for species $1$), and
$|\psi_{1,n}|^2+|\psi_{2,n}|^2$ in (a), (b), and (c), respectively.
The white dotted lines predict that, at the time $\lambda_{1,2}$
starts to vary. (d) The changing of $Rd_j(t)$ for this two species
with time. Its values refer to the symbiotic DB and are gained from
Eq. (\ref{eq11}), where $k_1=40$ and $k_2=42$. The solid and dotted
lines represent the  $Rd_j(t)$ of species $1$ and $2$, respectively.
Here, $\lambda_{1,2}=3$, $0.7$, $0.3$, and $0$ in the ranges from
the time steps $0-200$, $200-400$, $400-600$ and $600-800$,
respectively. $\lambda_{1,1}=0$ and $\lambda_{2,2}=9$ at all time.}
\end{figure}

\begin{figure}[!htp]
\includegraphics[width=8.5cm]{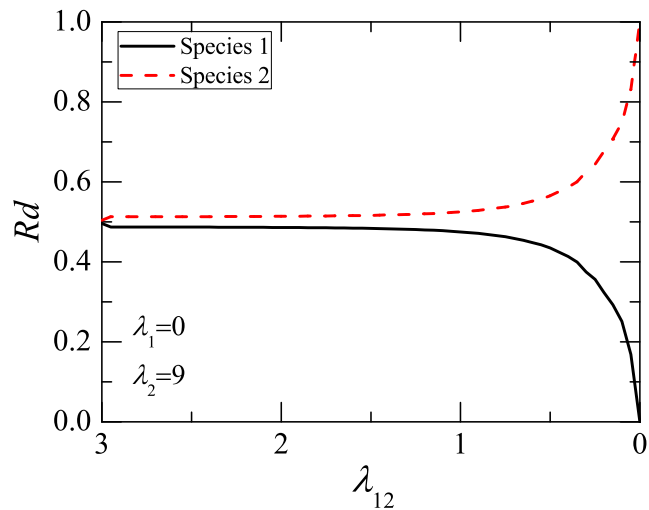}
\caption{\label{Fig7}  Correspondence relationship between the
$Rd_j(t)$ and $\lambda_{1,2}$. Its values refer to the symbiotic DB
and are gained from Eq. (\ref{eq11}), where $k_1=40$ and $k_2=42$.
The solid and dotted lines represent those of species $1$ and $2$,
respectively. Here $\lambda_{1,1}=0$ and $\lambda_{2,2}=9$.}
\end{figure}

\subsection{Selective distillation for controlling the $Rd_j(t)$ of two-species BECs.}
It  is interesting to control the $Rd_j(t)$ of two-species BECs in
open optical lattices with a stable  symbiotic DB by using selective
distillation (here $\gamma_1=\gamma_2=0.3$). As shown in the section
of results before,  when $\Lambda_{1,1}=0$ and
$\Lambda_{2,2}>\Lambda_{b}$, species $1$ will decay completely and
the DB can be created in species $2$ if $\Lambda_{1,2}=0$. However,
if $\Lambda_{1,2}$ is large, these two species can be co-localized
at the same location. That is, $\Lambda_{1,2}$ plays an important
role in the dynamics of the mixture, which means that one  can
control the $Rd_j(t)$ of the mixture of the two-species BECs by
manipulating the corresponding interspecies interaction.

Let us assume that this mixture is initially a stable symbiotic DB,
where the $Rd_j(t)$ of the two species are $50\%$, and this mixture
locates at the middle site in optical lattices with $M=81$. It can
be described as
\begin{eqnarray}\label{eq14}
\begin{array}{c}
\!\!A_{j,40}(t=0)=A_{j,42}(t=0)=0.1667,\;\;\;\;\;\;\;\;\;\;\;\;\;\;
\\ A_{j,41}(t\!=\!0)\!=\!\left(1\!
-\!A_{j,40}^2(t\!=\!0)\!-\!A_{j,42}^2(t\!=\!0)\right)^{1/2}, \\
\end{array}
\end{eqnarray}
where $j$ (=1,2) represents the two different species. By varying
$\lambda_{1,2}$ every $200$ time steps, the dynamics of the system
is shown in Fig. \ref{Fig6}. The white dotted lines label the time
that $\lambda_{1,2}$ starts to vary.  Fig. \ref{Fig6}(d) describes
the $Rd_j(t)$ change of species $1$ and $2$.

From Fig. \ref{Fig6}(a), one can clearly find that when $0<t<200$
and $\lambda_{1,2}=3$, the density of species $1$ is stable. The
$Rd_j(t)$ of species $1$ and $2$ near to be $50\%$, as shown in the
area from $t=0$ to $t=200$ in Fig. \ref{Fig6}(d). From  $t=200$ to
$t=400$, $\lambda_{1,2}$ turns to be $0.7$. The density of species
$1$ decreases suddenly to another stable value. Accordingly, the
$Rd_j(t)$ of species $1$ ($2$) decreases (increases) suddenly to
another stable value, as shown in the area from $t=200$ to $t=400$
in Fig. \ref{Fig6}(d). Similarly, from  $t=400$ to $t=600$,
$\lambda_{1,2}$ becomes much smaller to be $0.3$. During this time,
the density of species $1$ continues to decrease, and the $Rd_j(t)$
of species $1$ ($2$) decreases (increases) suddenly to a different
stable value, as shown in the area from $t=400$ to $t=600$ in Fig.
\ref{Fig6}(d). At $t=600$, $\lambda_{1,2}$ decreases to zero, and
species $1$ will  decay completely. Accordingly, the $Rd_j(t)$ of
species $1$ decreases to $0$, and that of species $2$ increases to
$100\%$, as shown in the area from $t=600$ to $t=800$ in Fig.
\ref{Fig6}(d). In the entire process, species $2$ does not change
and is always localized, as shown in Fig. \ref{Fig6}(b). The sum
density of both species $1$ and $2$ has been presented in Fig.
\ref{Fig6}(c). It predicts that each $\lambda_{1,2}$ corresponds to
a specific $Rd_j(t)$ of this mixture. Consequently, the $Rd_j(t)$ of
the two-species BECs mixture can be controlled by adjusting
$\lambda_{1,2}$.

We systemically calculate numerically the correspondence
relationship between the $Rd_j(t)$ and $\lambda_{1,2}$, shown in
Fig. \ref{Fig7}. It is obtained in the similar way used in Fig.
\ref{Fig6}, where $\lambda_{1,2}$ varies  a time every $200$ time
steps. It is worth noting that we had calculated  $Rd_j(t)$ of the
DB for different values of $\lambda_2$, and the results show that
they have no obvious difference. It is obvious that there is a
selective distillation phenomenon in the dynamics of a stable
symbiotic DB of two-species BECs.

\section*{Methods}

To investigate the formation of two-species BECs in an open optical
lattice, we supplement the standard DNLSEs with a local dissipation
at the two edges of the lattice. They are given by
\begin{eqnarray}\label{eq6}
\!\!\!\!\!\!\!\!i\frac{\partial \psi _{1,n}} {\partial
t}\!&\!=\!&\left(\lambda_{1,1} \left|\psi_{1,n}\right|^{2}+
\lambda_{1,2}\left|\psi_{2,n}\right|^{2} \right)\psi_{1,n}\nonumber
\\&&-\frac{1}{2}\left(\psi _{1,n-1}+\psi _{1,n+1}\!\right)
-i\gamma_{1}\psi_{1,n}(\delta_{n,1}+\delta_{n,M}),
\end{eqnarray}
\begin{eqnarray}\label{eq7}
\!\!\!\!\!\!\!\!i\frac{\partial \psi _{2,n}} {\partial
t}\!&\!=\!&\left(\lambda_{2,2} \left|\psi_{2,n}\right|^{2}+
\lambda_{1,2}\left|\psi_{1,n}\right|^{2} \right)\psi_{2,n}\nonumber
\\&&-\frac{1}{2}\!\left(\psi _{2,n-1}+\psi _{2,n+1}\!\right)
-i\gamma_{2}\psi_{2,n}(\delta_{n,1}+\delta_{n,M}),
\end{eqnarray}
where $\delta_{n,1}$ and $\delta_{n,M}$ are delta functions.
$\gamma_{j}$ describes the atom loss from the boundary of the
optical lattices. The optical lattices with leaking edges can be
realized experimentally by separating continuous microwave or Raman
lasers, where  $\gamma_{j}$ can be estimated within a mean-field
approximation\cite{{RLivi2006},{RFranzosi2007}}.

\section*{Conclusion}

We have numerically investigated the formation of DBs in two-species
BECs by DNLSEs in open optical lattices, and found that there is a
selective distillation phenomenon in the mixture of two-species
BECs. The coupling of intra- and interspecies interaction can lead
to the existence of pure single-species DBs and  symbiotic  DBs
(i.e., the two single-species DBs localized together in the same
sites), whose formation can be controlled by varying their
interactions in the two-species BECs. In this way, one can
selectively distil one species from the mixture of two-species BECs,
including the mixture of initial condition selected randomly, that
of moving symbiotic DBs, and that of a stable symbiotic DB,  and can
even control the dominant specie fraction by adjusting the
interspecies interaction in optical lattices. Maybe our selective
distillation of ultracold atomic gas is useful in quantum
information storage and quantum information processing based on
multi-species atoms.

\bigskip
\textbf{Acknowledgments:}  This work is supported by the National
Natural Science Foundation of China under Grant Nos.  11174040,
11475021, 11474026, and 11474027, and the Fundamental Research Funds
for the Central Universities under Grant No. 2015KJJCA01.

\bigskip
\textbf{Author contributions:}  X.D., J.X. and F.G. completed the
calculation and prepared the figures. X.D., M.Z., G.J. and F. G.
wrote the main manuscript text. F.G. supervised the whole project.
All authors reviewed the manuscript.


\end{document}